\documentclass[pra,twocolumn,superscriptaddress]{revtex4-1}
\usepackage{float,graphicx,multirow, amsmath,color,dsfont}
\usepackage{amsmath,bm}
\usepackage{amssymb,mathrsfs,dsfont}
\usepackage{amssymb,mathbbol}
\usepackage{amsfonts}
\usepackage{mathrsfs}
\usepackage{color,hyperref}
\definecolor{darkblue}{rgb}{0.0,0.0,0.3}
\hypersetup{colorlinks,breaklinks,
linkcolor=darkblue,urlcolor=darkblue,
anchorcolor=darkblue,citecolor=
darkblue,pdfauthor=Chandan, pdftitle=Chandan_Gaussian }
\newcommand{\etal}{\textit{et al}.\,}
\newcommand{\ie}{\textit{i}.\textit{e}.}

\newcommand\doubleRule{\toprule}
\begin{document}
\title{Optimal characterization of Gaussian channels using
photon-number-resolving detectors}
\author{Chandan Kumar}
\email{chandankumar@iisermohali.ac.in}
\affiliation{Department of Physical Sciences, Indian
Institute of Science Education and Research (IISER)
Mohali, Sector 81 SAS Nagar, Manauli PO 140306 Punjab India.}
\author{Ritabrata Sengupta}
\email{rb@iiserbpr.ac.in}
\affiliation{Department of Mathematical Sciences, Indian
Institute of Science Education and Research (IISER)
Berhmampur, Transit Campus, Government ITI, Berhampur  760
010, Odisha, India.}
\author{Arvind}
\email{arvind@iisermohali.ac.in}
\affiliation{Department of Physical Sciences, Indian
Institute of Science Education and Research (IISER)
Mohali, Sector 81 SAS Nagar, Manauli PO 140306 Punjab India.}
\begin{abstract}
We present optimal schemes, based on photon number
measurements, for Gaussian state tomography and for Gaussian
process tomography.  An $n$-mode Gaussian state is completely
specified by $2 n^2+3n$ parameters. Our scheme requires
exactly $2 n^2+3n$ distinct photon number measurements to
tomograph the state and is therefore optimal. Further, we
describe an optimal scheme to characterize Gaussian
processes by using coherent state probes and photon number
measurements.  With much recent progress in photon number measurement
experimental techniques, we hope that our scheme will be
useful in various quantum information processing protocols
including entanglement detection, quantum computation,
quantum key distribution and quantum teleportation. This
work builds upon the work of Parthasarathy \etal
[\href{http://dx.doi.org/10.1142/S021902571550023X}{Infin.
Dimens. Anal. Quantum Probab. Relat. Top., 18(4): 1550023,
21, 2015}]. 
\end{abstract}
\maketitle
%%%%%%%%%%%%%%%%%%%%%%%%%%%%%%%%%%
\section{Introduction}
\label{intro}
Continuous variable (CV) systems are ubiquitous in quantum
information and communication protocols.  Most of the CV
quantum information protocols are based on Gaussian states
as they are easy to prepare, manipulate and
measure~\cite{weedbrook-rmp-2012, adesso-2014}. One of the central
tasks in quantum information processing is the estimation of
quantum states which is formally called quantum state
tomography
(QST)~\cite{james-pra-2001,paris-2011,lvovsky-rmp}.
Generally, homodyne and heterodyne measurements are employed
in CV QST, which measure quadrature operators of a given
state~\cite{yuen-1982,yuen-83,vogel-pra-1989}.  However,
with the recent development of experimental techniques in
photon-number-resolving-detectors
(PNRD)~\cite{silberhorn-prl-2016,josef-prl-2019}, the
possibility of carrying out QST via photon number
measurements has opened up.  Cerf \etal devised a scheme
using beam splitters and on-off detectors, where one can
obtain the trace and determinant of the covariance
matrix of a Gaussian
state~\cite{cerf-prl-2004,cerf-pra-2004}. 
In a similar endeavor, Parthasarathy \etal have
developed a theoretical scheme to determine the Gaussian
state by estimating its mean and covariance
matrix~\cite{rb-2015}.

Another important task in quantum information processing is
quantum process tomography (QPT), where we wish to
characterize quantum processes which in general are
completely positive maps.  For CV systems, theoretical as
well as experimental studies for QPT  have been  undertaken
by several
authors~\cite{lobino-science,rahimi-njp-2011,anis-2012,
wang-pra-2013,cooper-njp-2015,
connor-reports-2015,jarom-pra-2015,rezakhani-pra-2017,
filip-reports-2017,dowling-pra-2018}.  Lobino \etal used
coherent state probes along with homodyne measurements to
characterize quantum processes~\cite{lobino-science}.
Similarly, Ghalaii \etal  have developed a coherent state
based QPT scheme via the measurement of normally ordered
moments that are measured using homodyne
detection~\cite{rezakhani-pra-2017}.  In this direction,
Parthasarathy \etal have utilized QST schemes based on
photon number measurements for Gaussian states, to
characterize the Gaussian channel~\cite{rb-2015}.

In this paper, we simplify the scheme given by Parthasarathy
\etal~\cite{rb-2015} and  describe an optimal scheme which
involves a minimum number of measurements and utilizes smaller
number of optical elements for the QST of Gaussian states
based on PNRD. We employ this scheme to devise  an optimal
scheme for Gaussian channel characterization.
An $n$-mode Gaussian state is completely specified by
their $2n$ first moments and second order moments arranged
in the form of a covariance matrix which has $2 n^2+n$
parameters. Therefore, we require a total of $2 n^2 +3n$
parameters to completely determine an $n$-mode Gaussian
state. The QST based on photon number measurements is
optimal in the sense that we require exactly $2 n^2 + 3 n$  distinct
measurements to determine all the $2 n^2 + 3 n$ parameters
of the state. Next we deploy the QST scheme that we develop,
to estimate the output, with coherent state probes as inputs
for the Gaussian channel characterization.  An $n$-mode
Gaussian channel is described by a pair of $2n \times 2n$
real matrices $A$ and $B$ with $B=B^T \geq 0$ which satisfy
certain complete positivity and trace preserving
conditions~\cite{heinosaari-2010,
holevo-2012,parthasarathy-2015}.  The matrices $A$ and $B$
together can be described by a total of $ 6n^2+n$
parameters.  We show that we can characterize a Gaussian
quantum channel optimally, \ie, we require exactly $6n^2+n$
distinct measurements to determine all the $6n^2+n$ parameters of the
Gaussian channel.  We compare the variance of transformed
number operators arising in the aforementioned QST scheme
which provides an insight into the efficiency of the scheme.
Finally, we relate the variance of transformed number
operators to the variance of quadrature operators.
 In CV quantum key distribution (QKD) protocols, one
needs to send an intense local oscillator pulse for the
purpose of measurement, which in itself is an arduous task
and can give rise to security
loopholes~\cite{sarovar-prx-2015, bing-prx-2015}. Our scheme
based on PNRD does not require such an intense local oscillator
signal, and thus may turn out to be useful in CV-QKD
protocols.

The paper is organized as follows. In Sec.~\ref{cvsystem} we
give a detailed mathematical background about CV
systems. In Sec.~\ref{sec:gaussian} we provide our optimal
QST scheme  based on PNRD for Gaussian states. Thereafter,
the tomography of the Gaussian channel has been dealt with in
Sec.~\ref{sec:channel} while in Sec.~\ref{sec:variance} we
compare the variance of different transformed number
operators appearing in the state tomography scheme.  Finally
in Sec.~\ref{sec:conclusion} we draw conclusions from our
results and look at future aspects.
%%%%%%%%%%%%%%%%%%%%%%%%%%%%%%%%%%%%%%%%%%%
\section{CV system}
\label{cvsystem}
An $n$-mode continuous variable 
quantum system is represented by $n$
pairs of Hermitian quadrature operators $\hat{q}_i,
\hat{p}_i$ ($i=1\,,\dots, n$)
which can be arranged in a column
vector
as~\cite{arvind1995,Braunstein,adesso-2007,weedbrook-rmp-2012,adesso-2014}
\begin{equation}\label{eq:columreal}
\bm{\hat{ \xi}} =(\hat{ \xi}_i)= (\hat{q_{1}},\,
\hat{p_{1}} \dots, \hat{q_{n}}, 
\, \hat{p_{n}})^{T}, \quad i = 1,2, \dots ,2n.
\end{equation}
The bosonic commutation relation  between them 
in a compact form read as ($\hbar$=1)
\begin{equation}\label{eq:ccr}
[\hat{\xi}_i, \hat{\xi}_j] = i \Omega_{ij}, \quad (i,j=1,2,...,2n),
\end{equation}
where $\Omega$ is the 2$n$ $\times$ 2$n$ matrix given by
\begin{equation}
\Omega = \bigoplus_{k=1}^{n}\omega =  \begin{pmatrix}
\omega & & \\
& \ddots& \\
& & \omega
\end{pmatrix}, \quad \omega = \begin{pmatrix}
0& 1\\
-1&0 
\end{pmatrix}.
\end{equation}
The  field annihilation and creation 
operators $\hat{a}_i\, \text{and}\, {\hat{a}_i}
^{\dagger}$ ($i=1,2,\, \dots\, ,n$) 
are related to the quadrature operators as 
\begin{equation}\label{realtocom}
\hat{a}_i=   \frac{1}{\sqrt{2}}(\hat{q}_i+i\hat{p}_i),
\quad  \hat{a}^{\dagger}_i= \frac{1}{\sqrt{2}}(\hat{q}_i-i\hat{p}_i).
\end{equation}

The number operator for the $i^{\text{th}}$ mode and total number operator
 for $n$-mode system can be expressed as
\begin{subequations}
\begin{align}\label{eq:generalcal}
\hat{N_i} = &\hat{a_i}^{\dagger}\hat{a_i} =
\frac{1}{2}\left( \hat{q_i}^2+\hat{p_i}^2 -1 \right), \\
\hat{N} = &\sum_{i=1}^{n}\hat{N_i}.
\end{align}
\end{subequations}
The state space known as Hilbert space $\mathcal{H}_i$  
for $i^\text{th}$ mode is  spanned
by the eigen vectors
$\vert n_i \rangle, \quad \{n_i=0,\,1, \dots ,\infty \} $ of
$N_i=a_i^{\dagger} a_i$. 
The combined Hilbert space $\mathcal{H}^{\otimes n} = 
\otimes_{i=1}^{n}\mathcal{H}_i$
of the $n$-mode state is spanned by the product basis
vector 
$ \vert n_1\dots n_i \dots n_n\rangle$ with $\{n_1,\, 
\dots\,, n_i,\, \dots\,, n_n=0,\, 1, \dots ,\infty \} $.
The numbers $n_i$
correspond to photon number in the $i^{\text{th}}$ mode.
The irreducible action of the field operators $\hat{a}_i$ and
$\hat{a}^{\dagger}_i$  
on $\mathcal{H}_i$ is dictated by 
the commutation
relation Eq.~(\ref{eq:ccr}) and is given by
\begin{equation}
\begin{aligned}
\hat{a_i}|n_i\rangle =& \sqrt{n_i}|n_i-1\rangle \quad n_i
\geq 1,
\quad\hat{a_i}|0\rangle = 0,\\
\hat{a_i}^{\dagger}|n_i\rangle = &\sqrt{n_i+1}|n_i+1\rangle
\quad n_i \geq 0.
\end{aligned}
\end{equation}
We define displacement operator acting on the $i^{\text{th}}$ mode
and the corresponding coherent states as:
\begin{eqnarray}
\hat{D}_i(q_i,p_i) &=& e^{i(p_i\hat{q}_i-q_i
\hat{p}_i)},\nonumber \\
\vert q_i, p_i\rangle_i &=&  \hat{D}_i(q_i,p_i)|0\rangle_i.
\end{eqnarray}
 Here $q_i$ and $p_i$ 
correspond to displacement along $\hat{q}$ and $\hat{p}$-quadrature of the $i^{\text{th}}$ mode.
%%%%%%%%%%%%%%%%%%%%%%%%%%%%%%%%%%%%%%%%%%%%%%%%%%%%%%%%%%%%%
\subsection{Symplectic transformations} 
The group $Sp(2n,\,\mathcal{R})$ is  defined as 
the group of linear homogeneous transformations $S$ specified by real 
$2n \times 2n$ matrices $S$ acting on the quadrature
variables and preserving the 
the  canonical commutation relation
Eq.~(\ref{eq:ccr}):
\begin{equation} 
\hat{\xi}_i \rightarrow
\hat{\xi}_i^{\prime} = S_{ij}\hat{\xi}_{j}, \quad\quad
S\Omega S^T = \Omega.  
\end{equation} 
The unitary representation of this
group turns out to be infinite dimensional where we have
$\mathcal{U}(S)$ for each $S \in Sp(2n,\, \mathcal{R})$
acting on a Hilbert space and is known as the metaplectic
representation.  These unitary transformations are generated
by Hamiltonian which are  quadratic functions of quadrature
and field operators.  Further, any symplectic matrix $S$
$\in$ $Sp(2n, \,\mathcal{R})$ can be decomposed as
\begin{equation}
S =  P \cdot T,
\end{equation}
$P$ $\in$ $\Pi(n)$ is a subset of  $Sp(2n,\, \mathcal{R})$
defined as
\begin{equation}
\Pi(n) = \{ S \in Sp(2n, \,\mathcal{R})\,|\,S^T =S,\,\,
S>0\},
\end{equation}
and $T$ is an element of $K(X,Y)$, the maximal compact
subgroup of $Sp(2n,\, \mathcal{R})$ which is isomorphic to
the unitary group $U(n)=X+iY$ in $n$-dimensions.  The action
of $U(n)$ transformation on the annihilation and creation
operators is given as
\begin{equation}
\label{unitary}
\bm{ \hat{a}} \rightarrow U \bm{ \hat{a}}, \quad 
\bm{ \hat{a}^{\dagger}} \rightarrow U^{*}\bm{ \hat{a}^{\dagger}},
\end{equation}
where $\bm{\hat{a}} =
(\hat{a}_1,\hat{a}_2,\dots,\hat{a}_n)^T$ and
$\bm{\hat{a}^{\dagger} }=
(\hat{a}_1^{\dagger},\hat{a}_2^{\dagger},\dots,\hat{a}_n^{\dagger})^T$.
The $2n\times 2n$ dimensional symplectic transformation
matrix $K(X,Y)$ acting on the Hermitian quadrature operators
can be easily obtained using Eqs.~(\ref{realtocom}) and
(\ref{unitary}).

Now we write 
three basic symplectic operations which will be used later.
%%%%%%%%%%%%%%%%%%%%%%%%%%%%%%%%%%%
\par
\noindent{\bf Phase change operation\,:}
The symplectic transformation for phase change operation 
acting on the quadrature operators $\hat{q}_i$, $\hat{p}_i$ is
\begin{equation}
R_i(\phi) = \begin{pmatrix}
\cos \phi & \sin \phi\\
-\sin \phi & \cos \phi
\end{pmatrix}.
\end{equation}
This operation corresponds 
to $U(1)$ subgroup of $Sp(2, \mathcal{R})$, its metaplectic
representation is generated by 
the Hamiltonian of the form 
$H =\hat{a}^{\dagger} _i\hat{a}_i$ and its action on 
the annihilation operator is
$\hat{a_i}\rightarrow e^{-i \phi} \hat{a}_i$.
\par
\noindent
{\bf Single mode squeezing operation\,:}
Symplectic transformation for the single mode squeezing operator
acting on quadrature operators $\hat{q}_i$ and $\hat{p}_i$
is written as
\begin{equation}
S_i(r) = \begin{pmatrix}
e^{-r} & 0 \\
0 & e^{r}
\end{pmatrix}.
\end{equation}
\par
\noindent
{\bf Beam splitter operation\,:}
For two-mode systems with 
quadrature operators
$  \hat{\xi} = (\hat{q}_{i}, \,\hat{p}_{i},\, \hat{q}_{j},\,
\hat{p}_{j})^{T}$ the
beam splitter transformation
$B_{ij}(\theta)$
acts as follows
\begin{equation}\label{beamsplitter}
B_{ij}(\theta) = \begin{pmatrix}
\cos \theta \,\mathbb{1}_2& \sin \theta \,\mathbb{1}_2 \\
-\sin \theta \,\mathbb{1}_2& \cos \theta \,\mathbb{1}_2
\end{pmatrix},
\end{equation}
where $\mathbb{1}_2$ represents $2 \times 2$ identity matrix
and transmittivity is specified  through  $\theta$ via 
the relation $\tau = \cos ^2 \theta$.  For a balanced
(50:50)
beam splitter, $\theta = \pi/4$.  
All the three operations above are generated by quadratic
Hamiltonians. It turns out that while phase change and beam
splitter operations are  compact and are generated by a photon
number conserving Hamiltonian, squeezing operations are
non-compact and are generated by a photon number non-conserving
Hamiltonian.

\subsection{Phase space description}
For a density
operator $\hat{\rho}$ of a quantum system the corresponding
Wigner distribution is defined as
\begin{equation}\label{eq:wigreal}
W(\bm{\xi}) = \frac{1}{{(2 \pi)}^{n}}\int \mathrm{d}^n \bm{q'}\, \left\langle
\bm{q}-\frac{1}{2}
\bm{q}^{\prime}\right| \hat{\rho} \left|\bm{q}+\frac{1}{2}\bm{\bm{q}^{\prime}}
\right\rangle \exp(i \bm{q^{\prime T}}\cdot \bm{p}),
\end{equation}
where
$\bm{\xi} = (q_{1}, p_{1},\dots, q_{n},p_{n})^{T}$,
$\bm{q^{\prime}} \in \mathcal{R}^{n}$ and $\bm{q} = (q_1,
q_2, \dots, q_n)^T$, 
$\bm{p} = (p_1, p_2, \dots, p_n)^T $.
Therefore, $W(\bm{\xi})$ depends upon 
$2n$ real phase space variables.

For an $n$ mode system,
the first order moments are defined as
\begin{equation}
\bm{d} = \langle \bm{\hat{\xi}} \rangle =
\text{Tr}[\hat{\rho}\bm{\hat{\xi}}],
\end{equation}
and the second order moments are best represented by the
real symmetric $2n\times2n$ covariance matrix defined as
\begin{equation}\label{eq:cov}
V = (V_{ij})=\frac{1}{2}\langle \{\Delta \hat{\xi}_i,\Delta
\hat{\xi}_j\} \rangle,
\end{equation}
where $\Delta \hat{\xi}_i = \hat{\xi}_i-\langle \hat{\xi}_i
\rangle$, and $\{\,, \, \}$ denotes anti-commutator.
The number of independent real parameters required to
specify the  covariance matrix is $n(2n+1)$.
The uncertainty principle in
terms of covariance matrix reads
$V+\frac{i}{2}\Omega \geq 0$ which
implies that the covariance matrix is positive
definite \ie, $V>0$.

A state is called a Gaussian state if the corresponding
Wigner distribution is a  Gaussian.  Gaussian states are
completely determined by their first and second order
moments and thus we require a total of $2n+ n(2n+1) = 2 n^2
+3n$ parameters to completely determine an $n$-mode Gaussian
state.
For the special case of Gaussian states, Eq.~(\ref{eq:wigreal}) 
can be written as~\cite{weedbrook-rmp-2012}
\begin{equation}\label{eq:wignercovariance}
W(\bm{\xi}) = \frac{\exp[-(1/2)(\bm{\xi}-\bm{d})^TV^{-1}
	(\bm{\xi}-\bm{d})]}{(2 \pi)^n \sqrt{\text{det}V}},
\end{equation}
where $V$ is the covariance matrix and  $\bm{d}$ 
denotes the displacement of the Gaussian state in phase
space.

We now compute averages of a few quantities that will be
required later, using the phase space representation. 
\begin{equation}\label{symmetric}
 \hat{N}  = \sum_{j=1}^{n}\hat{N_i} =
\frac{1}{2}\sum_{j=1}^{n}\left( \hat{q_i}^2+\hat{p_i}^2 -1 \right)
\end{equation}
is symmetrically ordered in $\hat{q}$ and $\hat{p}$
operators, therefore, 
average number of photons $\langle \hat{N} \rangle$ for an $n$-mode 
Gaussian state can be readily computed using the Wigner
distribution as follows~\cite{rb-2015, manko-pra-1994}:
\begin{eqnarray}
\langle \hat{N} \rangle  &=&\frac{1}{2}\sum_{j=1}^{n}\int
d^{2n} \bm{\xi}
 \left( q_i^2+p_i^2 -1 \right)  W(\bm{\xi}),\nonumber \\
&=&\frac{1}{2} \left[  \text{Tr}\left(
V-\frac{1}{2}\mathbb{1}_{2n}\right)+||\bm{d}||^2\right].
\label{avnumber}
\end{eqnarray}
Under a unitary transformation, while quantum states
transform in Schr\"odinger representation as
$\rho \rightarrow \,\mathcal{U} \rho
\,\mathcal{U}^{\dagger}$, in Heisenberg representation
the number operator transforms as, $\hat{N}
\rightarrow \,\mathcal{U}^{\dagger} \hat{N}
\,\mathcal{U}$.
Specifically for a phase space displacement $D(\bm{r})$, we have
\begin{equation}\label{disnumber1}
\langle \hat{D}(\bm{r})^\dagger \hat{N} \hat{D}(\bm{r})
\rangle =\frac{1}{2} \left[  \text{Tr}\left(
V-\frac{1}{2}\mathbb{1}_{2n}\right)+||\bm{d}+\bm{r}||^2\right],
\end{equation}
which simplifies by using  Eq.~(\ref{avnumber}) to
\begin{equation}\label{diffdis}
\langle \hat{D}(\bm{r})^\dagger \hat{N} \hat{D}(\bm{r}) \rangle
-\langle \hat{N} \rangle = \frac{1}{2}  \left( ||\bm{d}+\bm{r}||^2-||\bm{d}||^2 \right).
\end{equation}
For a homogeneous symplectic transformation $S$, 
the density operator follows the metaplectic representation $\mathcal{U}(S)$
as
$\rho \rightarrow \,\mathcal{U}(S) \rho
\,\mathcal{U}(S)^{\dagger}$.
The corresponding transformation of the displacement vector $\bm{d}$
and covariance matrix $V$ is given by~\cite{arvind1995}
\begin{equation}\label{transformation} 
\bm{d}\rightarrow S \bm{d},\quad  \text{and}\quad V\rightarrow SVS^T.
\end{equation}
Thus, we can easily evaluate the average of the number operator
after the state has undergone a  metaplectic transformation
using the Eqs.~(\ref{avnumber})~\&~(\ref{transformation}) as
\begin{equation}
\begin{aligned}
\langle \hat{\mathcal{U}}(S)^\dagger \hat{N}
\hat{\mathcal{U}}(S) \rangle =
\frac{1}{2} \text{Tr}\left( VS^T
S-\frac{1}{2}\mathbb{1}_{2n}\right)+\frac{1}{2} \bm{d}^T S^T
S  \bm{d}.
\end{aligned}
\end{equation}
Therefore,
\begin{equation}\label{diffsymplectic}
\begin{aligned}
\langle \hat{\mathcal{U}}(S)^\dagger \hat{N}
\hat{\mathcal{U}}(S) \rangle-\langle \hat{N} \rangle &=
\frac{1}{2}   \text{Tr}\left[ V(S^T S-\mathbb{1}_{2n})\right]\\
&+\frac{1}{2} \bm{d}^T (S^T S-\mathbb{1}_{2n})  \bm{d}.
\end{aligned}
\end{equation}
More mathematical details are available in~\cite{arvind1995}.
%%%%%%%%%%%%%%%%%%%%%%%%%%%%%%%%%%%%%%%%%%%%%%%%%%%%%%%%%%%%%%%%%
\section{Estimation of Gaussian states using photon number
measurements}
\label{sec:gaussian}
%%%%%%%%%%%%%%%%%%%%%%%%%
\begin{figure}[htbp]
\includegraphics[scale=1]{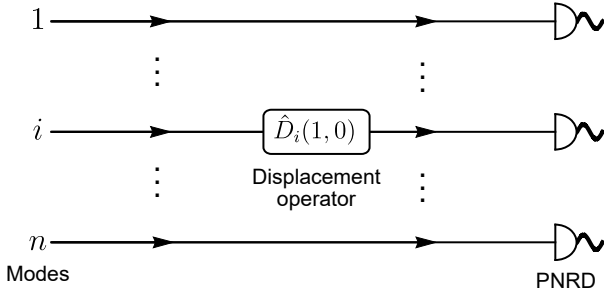} 
\caption{To estimate the mean of an $n$-mode Gaussian state,
the state is displaced along one of the $2n$ phase space
variables before performing photon number measurement on
each of the modes.  In the figure,  displacement gate
$\hat{D}_i(1,0)$ is applied on the state which displaces the
$\hat{q}$-quadrature of the $i^{\text{th}}$ mode by an unit
amount.}
\label{fig:disgate} 
\end{figure}
%%%%%%%%%%%%%%%%%%%%%%%%%%%%%%%%%%%
In this section, we present a variant of the scheme
developed in~\cite{rb-2015} where the authors have devised a
scheme to estimate the mean and covariance matrix of
Gaussian state using PNRD. In our scheme which is optimal
and uses minimum optical elements, photon number measurement
is performed on the original Gaussian state as well as
transformed Gaussian state. These transformations or gates
consist of displacement, phase rotation, single mode
squeezing and beam splitter operation denoted by
$\hat{D}_i(q,p)$, $\mathcal{U}(R_i(\theta))$,
$\mathcal{U}(S_i(r))$, and $\mathcal{U}  (B_{ij}(\theta))$,
respectively.
%%%%%%%%%%%%%%%%%%%%%%%%%%%%%%%%%%%%%%%%%%%%%%%%%%%%%%
\subsection{Mean estimation}\label{sec:dis}
We first perform photon number measurement on the original
$n$-mode Gaussian state giving us $\langle \hat{N} \rangle$.
Then we consider two different photon number measurements
after displacing one of the quadratures $\hat{q_i}$ or
$\hat{p_i}$ of the $i^{\text{th}}$ mode by an unit amount
giving us $\langle \hat{D}_i(1,0)^\dagger \hat{N}
\hat{D}_i(1,0) \rangle$ and $\hat{D}_i(0,1)^\dagger \hat{N}
\hat{D}_i(0,1) \rangle$.
(Figure~\ref{fig:disgate} depicts
displacement gate
$\hat{D}_i(1,0)$ acting on the $i^{\text{th}}$ mode
of the state.)  
We therefore have
by using
Eq.~(\ref{diffdis}):
\begin{eqnarray}
\langle \hat{D}_i(1,0)^\dagger \hat{N} \hat{D}_i(1,0)
\rangle-\langle \hat{N} \rangle
&= &\frac{1}{2}  \left( 1 +2 d_{q_i} \right),\nonumber \\
\langle \hat{D}_i(0,1)^\dagger \hat{N} \hat{D}_i(0,1)
\rangle-\langle \hat{N} \rangle
&= &\frac{1}{2}  \left( 1 +2 d_{p_i} \right),
\end{eqnarray}
which can be rewritten as
\begin{eqnarray}
d_{q_i} &=& \langle \hat{D}_i(1,0)^\dagger \hat{N}
\hat{D}_i(1,0) \rangle-\langle \hat{N} \rangle
-\frac{1}{2}, \nonumber \\
d_{p_i} &=& \langle \hat{D}_i(0,1)^\dagger \hat{N}
\hat{D}_i(0,1) \rangle-\langle \hat{N} \rangle
-\frac{1}{2}.
\label{meaneq}
\end{eqnarray}
Thus, we can obtain the mean values of $\hat{q}_i$ and
$\hat{p}_i$-quadratures once the values of $\langle
\hat{D}_i(1,0)^\dagger \hat{N} \hat{D}_i(1,0) \rangle$, $
\langle \hat{D}_i(0,1)^\dagger \hat{N} \hat{D}_i(0,1)
\rangle$, and $\langle \hat{N}\rangle$ have been obtained.
These estimations  involve measuring
averages and thus require us to repeat the measurement many
times.

Therefore, to obtain all the $2n$ elements of mean $\bm{d}$
of the Gaussian state, we need to perform $2n$ photon number
measurements after displacing the state by an unit amount
along $2n$ different phase spaces variables along with
photon number measurement on the original state.
We also note that $\text{Tr}(V)$ can be obtained using
Eq.~(\ref{avnumber}) once mean $\bm{d}$ of the Gaussian
state has been obtained.
\begin{equation}
\label{tracen}
\text{Tr}(V) = 2 \langle \hat{N} \rangle -||\bm{d}||^2+n.
\end{equation}
Thus, we are able to estimate $2n$ elements of mean
$\bm{d}$ of the Gaussian state and trace of the covariance
matrix $\text{Tr}(V)$ using a total of $2n+1$ photon number
measurements. 
%%%%%%%%%%%%%%%%%%%%%%%%%%%%%%%%%%%%%%%%%%%%%%%%%%%%%%%%
\subsection{Estimation of intra-mode covariance matrix}
\label{intramode}
%%%%%%%%%%%%%%%%%%%%%%%%%%%%%%%% 
 \begin{figure}[htbp]
\includegraphics[scale=1]{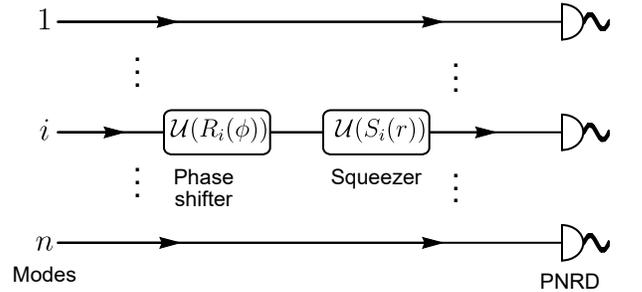} 
\caption{To estimate the intra-mode covariance matrix, that is, 
the covariance matrix of the individual modes, single mode 
symplectic transformations are applied on the state before 
performing photon number measurement on each of the modes. 
In the figure, phase shifter $\mathcal{U}(R_i(\phi)))$ followed
by a squeezer $\mathcal{U}(S_i(r))$ is applied on the $i^{\text{th}}$ 
mode of the state.}
\label{fig:symgate} 
\end{figure}

For convenience in representation, we express the covariance
matrix of the $n$-mode Gaussian state as follows: 
\begin{equation}
\label{nmodecov}
V=\begin{pmatrix}
V_{1,1}  & V_{1,2} &  \cdots    &  V_{1,n}       \\
V_{2,1}  &  \ddots & \ddots & \vdots  \\
\vdots &  \ddots & \ddots & V_{n-1,n}   \\
V_{n,1}  &  \cdots &  V_{n,n-1}      &V_{n,n}      
\end{pmatrix},
\end{equation}
where $V_{i,j}$ is a $2\times 2$ matrix. 
Further, we represent the mean and covariance matrix of the marginal 
state of mode $i$ (or intra-mode covariance matrix for mode $i$) as
\begin{equation}
d_i = \begin{pmatrix} d_{q_i}\\ d_{p_i}\end{pmatrix}, \quad
V_{i,i} = \begin{pmatrix} \sigma_{qq}&\sigma_{qp}\\
\sigma_{qp}&\sigma_{pp} \end{pmatrix}.
\end{equation}
To estimate the intra-mode covariance matrix, 
consider the single-mode symplectic gate $P_i(r,\phi)$
consisting of a squeezer and phase shifter acting on
the $i^{\text{th}}$ mode of the Gaussian state:
\begin{equation}
\label{phasesq}
P_i(r,\phi) = S_i(r) R_i(\phi) = 
\begin{pmatrix}
e^{-r} & 0 \\
0 & e^r
\end{pmatrix}
\begin{pmatrix}
\cos \phi & \sin \phi \\
-\sin \phi & \cos \phi
\end{pmatrix}.
\end{equation}
The schematic representation of $P_i(r,\phi)$ is shown in 
Fig.~\ref{fig:symgate}. 
When $P_i(r,\phi)$ acts on the $i^{\text{th}}$ mode 
of the Gaussian state, Eq.~(\ref{diffsymplectic}) reduces  to 
\begin{equation}
\label{phasesingle}
\begin{aligned}
\langle \hat{\mathcal{U}}(P_i)^\dagger \hat{N}
\hat{\mathcal{U}}(P_i) \rangle-\langle \hat{N} \rangle &=
\frac{1}{2}  \text{Tr}\left[ V_{i,i}(P_i^T P_i-\mathbb{1}_{2})\right]\\
&+\frac{1}{2}  d_i^T  (P_i^T P_i-\mathbb{1}_{2}) d_i.
\end{aligned}
\end{equation}
Here 
\begin{equation}
\begin{aligned}
P_i^T P_i=\begin{pmatrix}
e^{-2r}\cos^2 \phi+e^{2r}\sin^2 \phi&-\sinh 2r \sin 2\phi\\
-\sinh 2r \sin 2\phi&e^{-2r}\sin^2 \phi+e^{2r}\cos^2 \phi
\end{pmatrix}.
\end{aligned}
\end{equation}
For brevity, we assume 
\begin{equation}
\begin{aligned}
P_i^T P_i&-\mathbb{1}_{2}=\begin{pmatrix}
k_1&k_3\\
k_3&k_2
\end{pmatrix},
\end{aligned}
\end{equation}
and thus Eq.~(\ref{phasesingle}) simplifies as
\begin{equation}
\begin{aligned}
\langle \hat{\mathcal{U}}(P_i)^\dagger \hat{N} \hat{\mathcal{U}}(P_i)
 \rangle-&\langle \hat{N} \rangle =
\frac{1}{2} \bigg[  k_1 \sigma_{qq} +k_2 \sigma_{pp}+2k_3\sigma_{qp} \\
&+ k_1 {d_{q_i}}^2 +k_2 {d_{p_i}}^2+2 k_3 d_{q_i}d_{p_i}
\bigg].
\end{aligned}
\end{equation}
Rearranging the above equation, we obtain
\begin{equation}
\begin{aligned}\label{singlegate}
k_1 \sigma_{qq} +k_2 \sigma_{pp}+  2k_3\sigma_{qp}&=
2 \left( \langle \hat{\mathcal{U}}(P_i)^\dagger \hat{N}
\hat{\mathcal{U}}(P_i) \rangle
 -\langle \hat{N} \rangle \right)\\
& -(k_1 {d_{q_i}}^2 +k_2 {d_{p_i}}^2+2 k_3 d_{q_i}d_{p_i}).
\end{aligned}
\end{equation}
Since $d_{q_i}$ and $d_{p_i}$ have already been obtained in
Sec.~\ref{sec:dis}~(Eq.~(\ref{meaneq})), the above equation contains three
unknown parameters $\sigma_{qq}$, $\sigma_{pp}$, and
$\sigma_{qp}$. We can determine these three unknowns by
performing three distinct photon number measurement for
appropriate combinations of squeezing parameter $r$ and
phase rotation angle $\phi$, as follows:
\begin{itemize}
\item[(i)]
 For $e^{r} =\sqrt{2} $ and $\phi = 0$, we obtain
\begin{equation}\label{gate3}
-\frac{1}{2}\left( \sigma_{qq}-2\sigma_{pp}\right)  = c_1.
\end{equation}
\item[(ii)]
For $e^{r} =\sqrt{3} $ and $\phi = 0$, we obtain
\begin{equation}\label{gate4}
 -\frac{2}{3}\left(\sigma_{qq} -3 \sigma_{pp} \right) = c_2.
\end{equation}
\item[(iii)]
For $e^{r} =\sqrt{2} $ and $\phi = \pi/4$, we obtain
\begin{equation}\label{gate5}
\frac{1}{4} \left( \sigma_{qq} +\sigma_{pp} -6\sigma_{qp} \right) = c_3.
\end{equation}
\end{itemize}
Here $c_1$, $c_2$, and $c_3$ correspond to right-hand side
(RHS) of 
Eq.~(\ref{singlegate}) which can be easily determined once
the photon number measurements have been performed.
Equations~(\ref{gate3}) and~(\ref{gate4}) can be solved to
yield value of $\sigma_{qq}$ and $\sigma_{pp}$ which can be
put in Eq.~(\ref{gate5}) to obtain value of $\sigma_{qp}$.
Thus $V_{i,i}$ can be completely determined by performing
three photon number measurements after applying the three
distinct single mode symplectic gates
Eqs.~(\ref{gate3})-(\ref{gate5}). To determine all $V_{i,i}$
($1\leq i \leq n-1 $), we require $3(n-1)$ measurements. For
$V_{n,n}$, we need to determine 
$\sigma_{qp}$ and 
one of $\sigma_{qq}$ or
$\sigma_{pp}$, as $\text{Tr}(V)$ is already known.
Thus, a total of $3(n-1)+2=3n-1$ distinct
photon number measurements are required to determine all the
parameters of the intra-mode covariance matrix of a Gaussian
state.
%%%%%%%%%%%%%%%%%%%%%%%%%%%%%%%%%%%%%%%%%%%%%%%%%%%%%%%%%%%
\subsection{Estimation of inter-mode correlations matrix}
\begin{figure}[htbp]
\includegraphics[scale=1]{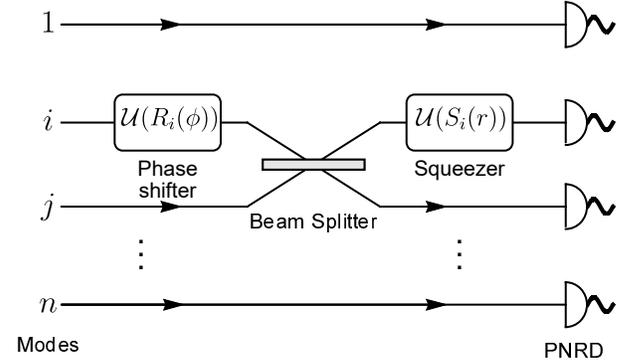} 
\caption{To estimate the inter-mode correlations matrix, we
apply a two-mode symplectic gate on the state before
performing photon number measurement on each of the modes.
As shown in the figure, first a  phase shifter
$\mathcal{U}(R_i(\phi)))$ is applied on the $i^{\text{th}}$
mode of the state. This is followed by a balanced beam
splitter $\mathcal{U}(B_{ij}(\frac{\pi}{4})))$ acting on $i$
$j$ modes and finally a squeezer $\mathcal{U}(S_i(r))$ is
applied on the $i^{\text{th}}$ mode of the state.}
\label{fig:twomode} 
\end{figure}
To estimate the inter-mode correlations matrix, we perform
two-mode symplectic operations on the Gaussian state before
measuring photon number distribution.  We write the
covariance matrix of the reduced state of the $i$ $j$ mode
in accord with Eq.~(\ref{nmodecov}) as
\begin{equation}
\begin{pmatrix} V_{i,i}& V_{i,j}\\V_{i,j}^T&V_{j,j}\end{pmatrix}.
\end{equation}
Here $i<j$ need not be successive modes. Since $V_{i,i}$ and
$V_{j,j}$ has 
already been determined in Sec.~\ref{intramode}, we need to
determine only $V_{i,j}$.
We further take the matrix elements of $V_{i,j}$ to be
\begin{equation}
\label{intermodecorrealtion}
V_{i,j} = \begin{pmatrix} \gamma_{qq}&\gamma_{qp}\\
\gamma_{pq}&\gamma_{pp} \end{pmatrix}.
\end{equation}
The two-mode symplectic gate is comprised of phase shifter
acting on the $i^{\text{th}}$ mode followed by a balanced
beam splitter acting on  modes $i$ $j$  and finally a
squeezer acting on mode $i$. 
We represent this mathematically as 
\begin{equation}
\label{twomodesym}
\begin{aligned}
Q_{ij}&(r,\phi) =(S_i(r) \oplus
\mathbb{1}_2)B_{ij}\left(\frac{\pi}{4}\right) (R_i(\phi)
\oplus \mathbb{1}_2), \\
&  = \begin{pmatrix}
S_i(r) & 0 \\
0& \mathbb{1}_2
\end{pmatrix}  
\frac{1}{\sqrt{2}}\begin{pmatrix}
\mathbb{1}_2& \mathbb{1}_2 \\
-\mathbb{1}_2& \mathbb{1}_2
\end{pmatrix}
\begin{pmatrix}
R_i(\phi) & 0 \\
0& \mathbb{1}_2
\end{pmatrix}. 
\end{aligned}
\end{equation}
The schematic representation of
$Q_{ij}(r,\phi)$  is illustrated in Fig.~\ref{fig:twomode}.
When the aforementioned gate $ Q_{ij}(r,\phi) $ acts on the
modes
$i$ $j$ of the Gaussian state,
Eq.~(\ref{diffsymplectic}) reduces  to 
\begin{equation}\label{twomodegate}
\begin{aligned}
\langle \hat{\mathcal{U}}(Q_{ij})^\dagger \hat{N}
&\hat{\mathcal{U}}(Q_{ij}) \rangle-\langle \hat{N} \rangle\\
& = \frac{1}{2}   \text{Tr}\bigg[ \begin{pmatrix} V_{i,i}&
V_{i,j}\\V_{i,j}^T&V_{j,j}\end{pmatrix}
\begin{pmatrix}K-\mathbb{1}_2& M\\M^T&L-\mathbb{1}_2\end{pmatrix}
\bigg]\\
 &+\frac{1}{2} \begin{pmatrix} d_{q_i}
\\d_{p_i}\\d_{q_j}\\d_{p_j} \end{pmatrix}^T
\begin{pmatrix}K-\mathbb{1}_2&
M\\M^T&L-\mathbb{1}_2\end{pmatrix}  \begin{pmatrix} d_{q_i}
\\d_{p_i}\\d_{q_j}\\d_{p_j} \end{pmatrix},
\end{aligned}
\end{equation}
where we have used
\begin{equation}
Q_{ij}^T Q_{ij}=  \begin{pmatrix}K& M\\M^T&L\end{pmatrix}.
\end{equation}
Using the following simplification for trace
\begin{equation}
\begin{aligned}
\text{Tr}&\bigg[ \begin{pmatrix} V_{i,i}& V_{i,j}\\V_{i,j}^T&V_{j,j}\end{pmatrix}
\begin{pmatrix}K-\mathbb{1}_2& M\\M^T&L-\mathbb{1}_2\end{pmatrix}
\bigg]\\
&=\text{Tr}\left[ V_{i,i}( K-\mathbb{1}_2)+  V_{j,j}
(L-\mathbb{1}_2 )\right] + 2\text{Tr}\left[ V_{i,j} M^T
\right],
\end{aligned}
\end{equation}
Eq.~(\ref{twomodegate}) can be rearranged as
\begin{eqnarray}
&&\text{Tr}\left[ V_{i,j} M^T \right]=\langle
\hat{\mathcal{U}}(Q_{ij})^\dagger \hat{N}
\hat{\mathcal{U}}(Q_{ij}) \rangle-\langle \hat{N}
\rangle \nonumber \\
&&\quad\quad  -\frac{1}{2}  \text{Tr}\left[ V_{i,i}(
K-\mathbb{1}_2)+
V_{j,j} (L-\mathbb{1}_2 )\right]\nonumber \\
 &&\quad\quad -\frac{1}{2} \begin{pmatrix} d_{q_i}
\\d_{p_i}\\d_{q_j}\\d_{p_j} \end{pmatrix}^T
\begin{pmatrix}K-\mathbb{1}_2&
M\\M^T&L-\mathbb{1}_2\end{pmatrix}  \begin{pmatrix} d_{q_i}
\\d_{p_i}\\d_{q_j}\\d_{p_j} \end{pmatrix}.
\label{doublegate}
\end{eqnarray}
Various terms appearing in the RHS of the above equation,
for instance $V_{i,i}$, $V_{j,j}$,
$d_{q_i}$, $d_{p_i}$, $d_{q_j}$, $d_{p_j}$ 
have already been determined.
Thus the four unknowns $\gamma_{qq}$, $\gamma_{pp}$,
$\gamma_{qp}$, $\gamma_{pq}$ appearing on the LHS of the
above equation can be determined by performing four
different photon number measurements for appropriate
combinations of squeezing parameter $r$ and phase rotation
angle $\phi$.  Further, LHS of Eq.~(\ref{doublegate})
can be expressed as following:
\begin{eqnarray}
&&\text{Tr}\left[V_{i,j} M^T \right] = 
\frac{1}{2}\bigg[ \left( e^{-2r}-1 \right) \cos \phi \, \gamma_{qq}
+\left( e^{2r}-1 \right) \cos \phi \, \gamma_{pp}\nonumber \\
&&\quad\quad\quad\quad+\left(1- e^{2r} \right) \sin \phi\, \gamma_{qp}
+\left( e^{-2r}-1 \right) \sin \phi \,\gamma_{pq}
\bigg].
\end{eqnarray}
We take these four different combinations of squeezing
parameter $r$ and phase rotation angle $\phi$ to determine
the four unknowns:

\begin{itemize}
\item[(i)]
 For $e^{r} =\sqrt{2} $ and $\phi = 0$, we obtain
\begin{equation}\label{twomodeg1}
-\frac{1}{4}\left( \gamma_{qq}-2\gamma_{pp}\right)  = d_1.
\end{equation}
\item[(ii)]
For $e^{r} =\sqrt{3} $ and $\phi = 0$, we obtain
\begin{equation}\label{twomodeg2}
-\frac{1}{3}\left(\gamma_{qq} -3 \gamma_{pp} \right) = d_2.
\end{equation}
\item[(iii)]
For $e^{r} =\sqrt{2} $ and $\phi = \pi/2$, we obtain
\begin{equation}\label{twomodeg3}
-\frac{1}{4}\left(2\gamma_{qp} + \gamma_{pq} \right) = d_3.
\end{equation}
\item[(iv)]
For $e^{r} =\sqrt{3} $ and $\phi = \pi/2$, we obtain
\begin{equation}\label{twomodeg4}
-\frac{1}{3}\left(3\gamma_{qp} + \gamma_{pq} \right) = d_4.
\end{equation}
\end{itemize}
Here $d_1$, $d_2$, $d_3$, and $d_4$ are the RHS of
Eq.~(\ref{doublegate}) which can be easily determined once
the photon number measurements have been performed.
Equations (\ref{twomodeg1}) and  (\ref{twomodeg2}) can be
solved to yield values of $\gamma_{qq}$ and $\gamma_{pp}$, 
whereas Eqs.~(\ref{twomodeg3}) and  (\ref{twomodeg4}) can be
solved to yield value of $\gamma_{qp}$ and $\gamma_{pq}$.
Thus, we have used four distinct measurements to determine
the four parameters of $V_{i,j}$. The inter-mode
correlations of the Gaussian states thus require $4\times
n(n-1)/2 = 2n(n-1)$ measurements.  So the total number of distinct
measurements required to determine all the $2 n^2 + 3 n$
parameters of the $n$-mode Gaussian state adds up to  $2 n^2
+ 3 n$.  The results are summarized in Table~\ref{table1}.
Thus, our tomography scheme for Gaussian state using photon
number measurement is optimal in the sense that we require
exactly the same number of distinct measurements as the
number of independent real parameters of the Gaussian state.
%%%%%%%%%%%%%%%%%%%%%
%{\nr Minimal nature to be brought out}
%%%%%%%%%%%%%%%%%%%%%%%%%%%%%%%%%%%%%%%
\begin{table}[ht!]
\caption{\label{table1}
Tomography of an $n$-mode Gaussian state by photon number
measurements}
\begin{tabular}{ p{2.5cm} p{2cm} p{2cm}  p{1.9cm} }
\doubleRule
Estimate type & Parameters number & Gaussian ~~~ Operations&
Measurement number \\
\doubleRule
Mean ($\bm{d}$)& $2n$& Displacement & $~~2n+1$\\
Intra-mode ~~~~~~~~~~ covariance ($V_{i,i}$)& $3n$& Phase
shifter, squeezer & $~~3n-1$\\
Inter-mode ~~~~~~~~~~~ correlations ($V_{i,j})$& $2n(n-1)$& 
Phase shifter, squeezer, beam splitter & $~~2n(n-1)$\\
\doubleRule
{\bf Total} &$\bm{2 n^2 + 3 n}$  & &$\bm{2
n^2 + 3 n}$ \\ \doubleRule
\end{tabular}
\end{table}
%%%%%%%%%%%%%%%%%%%%%%%%%%%%%%%%
\section{Characterization of Gaussian channels}
\label{sec:channel}
\begin{figure}[htbp]
\includegraphics[scale=1]{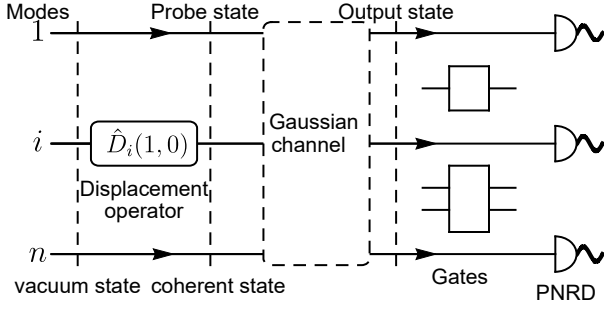} 
\caption{Scheme for a complete characterization of an
$n$-mode Gaussian channel. We send $2n$ distinct coherent state
probes through the channel and full or partial
state tomography is carried out on the output states.  In the
figure, displacement operator $\hat{D}_i(1,0)$ displaces the
$\hat{q}$-quadrature of the $i^{\text{th}}$  mode  by an
unit amount of an $n$-mode vacuum state to give one of the
required probe state. Single and two mode gate operations
involved in state tomography and described in
Section~\ref{sec:gaussian} are indicated as ``Gates''.
\label{fig:channel} }
\end{figure}
In this section, we move on to the characterization of a
Gaussian channel using coherent state
probes~\cite{lobino-science,rahimi-njp-2011,wang-pra-2013}
by employing the tomography techniques developed in
Section~\ref{sec:gaussian}.  Gaussian channels are defined
as those channels which transform Gaussian states into
Gaussian states~\cite{holevo-2012,parthasarathy-2015}.  An
$n$-mode Gaussian channel is specified by a pair of $2n
\times 2n$ real matrices $A$ and $B$ with $B=B^T \geq
0$~\cite{heinosaari-2010}.  The matrices $A$ and $B$ are
described by a total of $4 n^2 + 2n(2n+1)/2 = 6n^2+n$ real
parameters and satisfy complete positivity and trace
preserving condition
\begin{equation}
B+ i\Omega -i A \Omega A^T \geq 0.
\end{equation}
The action of the Gaussian channel on mean $\bm{d}$ and
covariance matrix $V$ of a Gaussian state is given by
\begin{equation}
\label{gtransform} 
\bm{d} \rightarrow A \bm{d}, \quad V \rightarrow
AVA^T+\frac{1}{2}B.
\end{equation}
Here again we follow the scheme proposed in \cite{rb-2015}. 
Schematic diagram of the scheme is shown in Fig.~\ref{fig:channel}.
We prepare $2n$  distinct coherent state probes by displacing $n$-mode
vacuum state by an unit amount along any of the $2n$
different phase space variables. These coherent state probes
are sent through the channel and full or partial state tomography
using photon number measurement is carried out on the output
states. The information about the output state parameters
enable us to characterize the Gaussian channel. Now we
describe the exact scheme in detail.  For convenience, we
define a $2n$ dimensional column vector as
\begin{equation}
\bm{e}_j = (0,0,\dots,1,\dots,0,0)^T,
\end{equation}
with $1$ present at the $j^{\text{th}}$ position.  First set
of $n$ coherent state probes are prepared by displacing
$n$-mode vacuum state ($\bm{d}=0$, $V=\mathbb{1}_{2n}/2$) by
an unit amount along  $n$ different $\hat{q}$-quadratures.
For instance, application of displacement operator
$\hat{D}_j(1,0)$ on the $j^{th}$ mode of the $n$-mode vacuum
state yields the coherent state
\begin{equation}
\begin{aligned}
|\bm{e}_{2 j-1}\rangle=\hat{D}_j(1,0) |\bm{0}\rangle,
\end{aligned}
\end{equation}
where $|\bm{0}\rangle$ denotes $n$-mode vacuum state. The
mean and covariance
matrix of the coherent state $|\bm{e}_{2 j-1}\rangle$ is given by 
\begin{equation}
\bm{d} = \bm{e}_{2 j-1}, \quad V= \frac{1}{2}\mathbb{1}_{2n}.
\end{equation}
This coherent state is sent through the Gaussian channel and
the mean and covariance matrix of the probe state transforms
according to Eq.~(\ref{gtransform}):
\begin{equation}
\label{outputstate}
\bm{d}_G=A\bm{e}_{2j-1}, \quad V_G=\frac{1}{2}(A A^T +B).
\end{equation}
Now we perform full state tomography on the output state
$\rho_G$ ($j=1$) which requires $2 n^2+3 n$ measurements.
This provides us the matrix  $A A^T +B$ and the first
column of  matrix $A$. For the rest $n-1$ probe states ($2
\leq j \leq n$), we measure only the mean of the output
state $\rho_G$  which enables us to determine all the odd
columns of matrix $A$.

However, as we noticed in Sec.~\ref{sec:dis}, we need to
perform $2n+1$ measurements to obtain the $2n$ elements of
mean vector $\bm{d}_G$. 
 This leads
to overshooting of the required number of measurements
compared to the number of channel parameters for the
complete characterization of Gaussian channel, which renders
the scheme non-optimal.  However, as we can see from
Eq.~(\ref{outputstate}),
 $\text{Tr}(V) =
\text{Tr}(A A^T +B)/2$  is same for all probe states as all
the output states have the same covariance
matrix and has already been obtained in
the process of tomography of the first output state ($j=1$).
Now we show how this fact can be exploited to obtain the
value of $\langle \hat{N} \rangle$ for the other coherent
state probes,  resulting in an optimal characterization
of the Gaussian channel.
We perform  $2n$ measurements after displacing the output
state $\rho_G$ corresponding to second coherent state probe
and obtain $2n$ equations as follows: 
\begin{equation}
\label{channelm}
\begin{aligned}
d_{q_i} =& \langle \hat{D}_i(1,0)^\dagger \hat{N} \hat{D}_i(1,0) 
\rangle-\langle \hat{N} \rangle - \frac{1}{2}, \quad 1 \leq i \leq n, \\
d_{p_i} =& \langle \hat{D}_i(0,1)^\dagger \hat{N} \hat{D}_i(0,1) 
\rangle-\langle \hat{N} \rangle - \frac{1}{2}, \quad 1 \leq i \leq n.
\end{aligned}
\end{equation}
We substitute $d_{q_i}$ and $d_{p_i}$ ($1 \leq i \leq n$) in
Eq.~(\ref{tracen}) and obtain a quadratic
equation in  $\langle \hat{N} \rangle$. After solving for
$\langle \hat{N} \rangle$, we put its value in
Eq.~(\ref{channelm}) to obtain $d_{q_i}$ and $d_{p_i}$ ($1
\leq i \leq n$).  
 Thus for other output
states $\rho_G$ ($2
\leq j \leq n$), only $2n$ measurements are required to
determine the mean vector $\bm{d}_G$ and no additional measurements are
required.

The other set of $n$ coherent state probes are prepared by
displacing $n$-mode vacuum state by an unit amount along
$n$ different $\hat{p}$-quadratures.  For instance,
application of displacement operator $\hat{D}_j(0,1)$ on the
$j^{th}$ mode of the $n$-mode vacuum state yields the
coherent state
\begin{equation}
\begin{aligned}
|\bm{e}_{2 j}\rangle=\hat{D}_j(0,1) |\bm{0}\rangle.
\end{aligned}
\end{equation}
The mean and covariance matrix of the coherent state
$|\bm{e}_{2 j}\rangle$ is given by 
\begin{equation}
\bm{d} = \bm{e}_{2 j}, \quad V= \frac{1}{2}\mathbb{1}_{2n}.
\end{equation}
This coherent state is sent through the Gaussian channel and
the mean and covariance matrix of the probe state transforms
according to Eq.~(\ref{gtransform}):
\begin{equation}
\bm{d}_G=A\bm{e}_{2j}, \quad V_G=\frac{1}{2}(A A^T +B).
\end{equation}
For all these $n$ output states $\rho_G$  ($1 \leq j \leq n$
), we measure only the mean which enables us to determine
all the even columns of matrix $A$. This information
completely specifies matrix $A$ as odd columns had already
been determined using the first set of $\hat{q}$-displaced
$n$ coherent state probes. This also enables us to obtain
matrix $B$ as matrix  $A A^T +B$ was already known from
the full state tomography on the first coherent state probe.
Thus, the total number of distinct measurements required sum up to
$6n^2+n$ as shown in Table~\ref{table2} which exactly
coincides with the parameters specifying a Gaussian channel.
In the scheme of
Parthasarathy \etal~\cite{rb-2015}, $2n-1$ additional
measurements were required which we do not need, 
leading to the optimality of our scheme. 
 We note that the scheme is optimal even  
when the coherent state probes have different mean values,
since $\text{Tr}(V)$ is same for all the output states even in
this case.
\begin{table}[h]
\caption{\label{table2}
Tomography of an $n$-mode Gaussian channel }
\begin{tabular}{ p{1.8cm} p{2.7cm}  c p{3cm}}
\doubleRule
Coherent state probe& Information ~~~~~~~ obtained&
&Measurement ~~~~~
number  \\
\doubleRule
$\hat{q}$-displaced& Odd columns of $A$ 
\& $(A A^T +B)$
&&  $2n^2+3n +(n-1) \times 2n$\\
$\hat{p}$-displaced & Even columns of $A$ && $n \times 2n$\\
\doubleRule
&{\bf Total} && $\bm{6n^2+n}$ \\ \doubleRule
\end{tabular}
	%	\end{ruledtabular}
\end{table}
%%%%%%%%%%%%%%%%%%%%%%%%%%%%%%%%%%%%%%%%%%%%%%%%%%%%%
\section{Variance in photon number measurements}
\label{sec:variance}
In this section, we analyze the variance of photon number
distribution of the original state and gate transformed
states which we used towards state and process estimation in
Sections~\ref{sec:gaussian}~\&~\ref{sec:channel}.  This
study will provide us with an idea of the quality of our
estimates of the Gaussian states and channels.

To evaluate the variance of photon number we note that the
square of the number operator can be easily put in
symmetrically ordered form as follows:
\begin{eqnarray}
&&\hat{N}^2 =\frac{1}{4}\sum_{i,j=1}^{n} \left(
\hat{q_i}^2+\hat{p_i}^2 -1 \right)
\left( \hat{q_j}^2+\hat{p_j}^2 -1 \right) \nonumber \\
&&\{\hat{N}^2\}_{\rm sym}=f(\hat{q},\hat{p})=
\frac{1}{4}\sum_{\substack{i,j=1\\ i\ne j}}^{n} \left(
\hat{q_i}^2+\hat{p_i}^2 -1 \right)
\left( \hat{q_j}^2+\hat{p_j}^2 -1 \right)\nonumber \\
&&\!\!\!\!+\frac{1}{4}\sum_{i=1}^{n}\! \bigg[ \hat{q_i}^4+\hat{p_i}^4
-2\hat{q_i}^2-2\hat{p_i}^2 \nonumber 
\!+\!\frac{1}{3}(\hat{q}_i^2\hat{p}_i^2+
\hat{q}_i\hat{p}_i\hat{q}_i\hat{p}_i+
\hat{q}_i\hat{p}_i^2\hat{q}_i)\!\bigg].
\nonumber \\
\end{eqnarray}
Thus
the average of $\hat{N}^2$ can be readily evaluated as
\begin{equation}
\begin{aligned}\label{av}
\langle \hat{N}^2 \rangle  =\int \mathrm{d}^{2n} \bm{\xi} \; f(q,p)
W(\bm{\xi}).
\end{aligned}
\end{equation}
Using the above equation and Eq.~(\ref{avnumber}), variance
of number operator can be written in an elegant form
as~\cite{manko-pra-1994,rb-2015,Pierobon-pra-2019}
\begin{equation}
\begin{aligned}\label{ab1}
\text{Var} (\hat{N} )  =&\langle \hat{N}^2 \rangle- \langle
\hat{N} \rangle^2\\
=&\frac{1}{2}   \text{Tr}\left[\left(
V-\frac{1}{2}\mathbb{1}_{2n}\right) \left(
V+\frac{1}{2}\mathbb{1}_{2n}\right) \right]+\bm{d}^T V \bm{d}.
\end{aligned}
\end{equation}
We first explore the mean and variance of photon number of
a single mode system to get some insights.  We consider a
single mode Gaussian state with mean $\bm{d} = (u,u)^T$ and
covariance matrix
\begin{equation}
\label{single-mode}
V (\beta)= \frac{1}{2}(2 \mathcal{N} +1)R(\beta)S(2s)R(\beta)^T,
\end{equation}
where $\mathcal{N}$ is the thermal noise parameter, $s$ is the 
squeezing and $\beta$ is the phase shift angle.
The mean and variance of the number operator for the above state reads
\begin{eqnarray}
&&\langle \hat{N} \rangle= \mathcal{N} \cosh 2s + \sinh^2 s +u^2,
\nonumber \\
&&\text{Var} (\hat{N}) =\left( \mathcal{N}+\frac{1}{2}
\right)^2 \cosh 4s 
-\frac{1}{4} \nonumber \\
&&
\quad\quad\,+2 u^2 \left( \mathcal{N}+\frac{1}{2} \right)
\left(\cosh 2s + \sin 2 \beta \sinh 2s \right).
\end{eqnarray}
Here both mean and variance depend on displacement $u$ and 
squeezing $s$ of the state. However, the mean photon number is 
independent of the phase shift angle $\beta$ while variance
of photon number depends on $\beta$.
 The variance of displaced number operator is given by
\begin{equation}
\label{vardis}
\begin{aligned}
&\text{Var}
\left(\hat{D}(\bm{r})^{\dagger}\hat{N}\hat{D}(\bm{r})\right
)  =
(\bm{d}+\bm{r})^T V  (\bm{d}+\bm{r})\\
&\quad\quad+\frac{1}{2}   \text{Tr}\left[\left(
V-\frac{1}{2}\mathbb{1}_{2n}\right) \left(
V+\frac{1}{2}\mathbb{1}_{2n}\right) \right].
\end{aligned}
\end{equation}
\begin{figure}[htbp]
\includegraphics[scale=1]{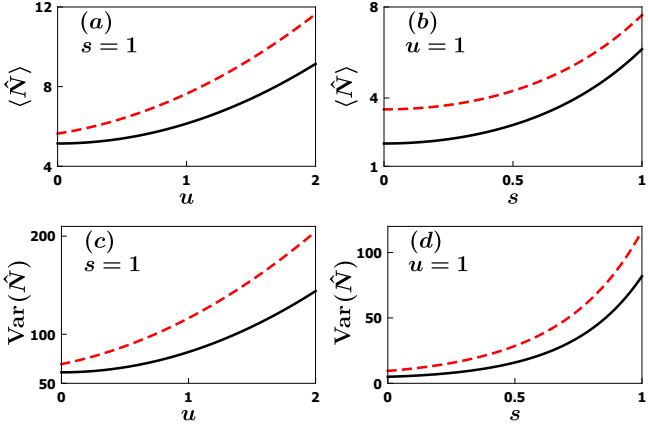} 
\caption{[Colour online] 
Single-mode squeezed coherent thermal
state~(\ref{single-mode}) has been plotted with parameters
$\beta = \pi/3$ and $\mathcal{N}=1$. For all four panels,
Black solid curve represents mean and variance of $\hat{N}$
while	Red dashed curve represents mean and variance of
$D^\dagger (1,0) \hat{N}D (1,0)$. $(a)$ Mean  photon number
as a function of  displacement $u$.  $(b)$ Mean  photon
number as a function of squeezing  $s$.  $(c)$ Variance of
photon number as a function of displacement $u$.  $(d)$
Variance of  photon number as a function of squeezing  $s$.
\label{fig:meanvar1} 
}
\end{figure}

In Fig.~\ref{fig:meanvar1}$(a)$, we plot $ \langle
\hat{N}\rangle$ and  $\langle D^\dagger (1,0) \hat{N}D
(1,0)\rangle$ as a function of displacement parameter $u$
for a single-mode squeezed coherent thermal
state~(\ref{single-mode}). We see that while $\langle
D^\dagger (1,0) \hat{N}D (1,0)\rangle$ is larger than
$\langle \hat{N}\rangle$, the mean values of both the
operators increases with the displacement parameter $u$.
Further,  $\langle D^\dagger (1,0) \hat{N}D
(1,0)\rangle$ equals  $\langle D^\dagger (0,1) \hat{N}D
(0,1)\rangle$ as can be seen from Eq.~(\ref{disnumber1}).
Similarly, Fig.~\ref{fig:meanvar1}$(b)$ shows that mean
values 
$ \langle \hat{N}\rangle$ and  $\langle D^\dagger (1,0) \hat{N}D
(1,0)\rangle$
increase with squeezing $s$.  We plot the variance of
the operators $ \hat{N}$ and  $D^\dagger (1,0) \hat{N}D (1,0)$
as a function of displacement parameter $u$ in
Fig.~\ref{fig:meanvar1}$(c)$. We see that while the variance
of operator $D^\dagger (1,0) \hat{N}D (1,0)$ is larger than
the variance of the operator $ \hat{N}$, variance of both the
operators increase with displacement parameter $u$.
Similarly, Fig.~\ref{fig:meanvar1}$(d)$ shows that variance of
the operators $ \hat{N}$ and  $D^\dagger (1,0) \hat{N}D
(1,0)$ increase with squeezing $s$.

\begin{figure}[htbp]
\includegraphics[scale=1]{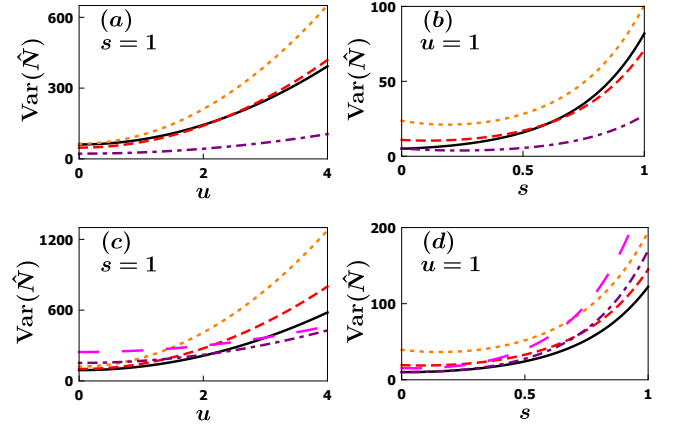} 
\caption{[Colour online] $(a)$ Variance of  photon number as
a function of  displacement $u$ for single-mode
squeezed thermal state~(\ref{single-mode}). 
$(b)$ Variance of  photon number as
a function of  squeezing $s$ for single-mode squeezed
thermal state~(\ref{single-mode}). 
For both panel $(a)$ and $(b)$, various curves correspond to
$\text{Var} (\mathcal{U}^\dagger(P)\hat{N}\mathcal{U}(P))$
with  $e^r=\sqrt{2}$, $\phi= 0$ (Red dashed),
$e^r=\sqrt{3}$, $\phi= 0$ (Orange dotted), $e^r=\sqrt{2}$, 
$\phi= \pi/4$ (Purple dot dashed), while Black solid curve
represents $\text{Var} (\hat{N})$, and parameter $\beta =\pi/3$. 
$(c)$ Variance of  photon number as a function of  displacement
$u$ for two-mode squeezed thermal state~(\ref{two-modestate}). 
$(d)$ Variance of  photon number as a function of  squeezing $s$
for two-mode squeezed thermal state~(\ref{two-modestate}). 
For both panel $(c)$ and $(d)$, various curves correspond to 
$\text{Var} (\mathcal{U}^\dagger(Q)\hat{N}\mathcal{U}(Q))$ with
$e^r=\sqrt{2}$, $\phi= 0$ (Red dashed), $e^r=\sqrt{3}$, $\phi= 0$
(Orange dotted), $e^r=\sqrt{2}$, $\phi= \pi/2$ (Purple dot dashed),
$e^r=\sqrt{3}$, $\phi= \pi/2$ (Magenta large dashed) while Black 
solid curve represents $\text{Var} (\hat{N})$. For all four panels,
thermal parameter has been taken as $\mathcal{N}=1$. } 
\label{fig:meanvar12} 
\end{figure}

The variance of photon number 
after a symplectic transformation $S$ of the  state reads
as:
\begin{equation}
\label{symvar} 
\begin{aligned} &\text{Var}
\big(\mathcal{U}(S)^{\dagger}\hat{N}\mathcal{U}(S))\big )
= \bm{d}^T V \bm{d}\\ &\quad\quad+\frac{1}{2}
\text{Tr}\left[\left(
SVS^T-\frac{1}{2}\mathbb{1}_{2n}\right) \left(
SVS^T+\frac{1}{2}\mathbb{1}_{2n}\right) \right].\\
\end{aligned} 
\end{equation}
Using this expression we first compare the variance of 
number operator under the action of $P_i(r,\phi)$
gate (Eqn.~(\ref{phasesq})) for different values of the
parameters $r$ and $\phi$. In Fig.~\ref{fig:meanvar12}($a$), we
plot the variance of different $P_i(r,\phi)$ gate
transformed number operators as a function of displacement
$u$ for single-mode squeezed coherent thermal
state~(\ref{single-mode}). We can see that the variance of
different $P_i(r,\phi)$ gate transformed number operators
increase  with displacement $u$. While the variance of
$\mathcal{U}^\dagger(P)\hat{N}\mathcal{U}(P)$ with
$e^r=\sqrt{2}$, $\phi= \pi/4$ is always lower than the
variance of $\hat{N}$ and variance of
$\mathcal{U}^\dagger(P)\hat{N}\mathcal{U}(P)$ with
$e^r=\sqrt{3}$, $\phi= 0$ is always higher than the variance
of $\hat{N}$, variance of
$\mathcal{U}^\dagger(P)\hat{N}\mathcal{U}(P)$ with
$e^r=\sqrt{2}$, $\phi= 0$ crosses over the variance of
$\hat{N}$ at a certain value of displacement $u$.  We show
the variance of the photon number as a function of squeezing
parameter $s$ in Fig.~\ref{fig:meanvar12}($b$). As we can see,
variance of different $P_i(r,\phi)$ gate transformed number
operators show a similar dependence on squeezing $s$ as that
of displacement $u$.

Now to compare the variance of photon number under the
action of two mode gates $Q_{ij}(r,\phi)$ (Eqn.
(\ref{twomodesym})), we
consider a two mode Gaussian state with mean $\bm{a} =
(u,u,u,u)^T$ and covariance matrix V
\begin{equation}\label{two-modestate}
V = B_{12}\left(\frac{\pi}{4}\right)[V(\beta_1) \oplus
V(\beta_2)]B_{12}\left(\frac{\pi}{4}\right)^T,
\end{equation}
where $V (\beta)$ is defined in Eq.~(\ref{single-mode}). We
use Eq.~(\ref{symvar}) to compute the variance of
$Q_{ij}(r,\phi)$ gate transformed number operator
corresponding to the above state.

In Fig.~\ref{fig:meanvar12}$(c)$, we plot the variance of
different $Q_{ij}(r,\phi)$ gate transformed number operators
as a function of displacement $u$ for two-mode squeezed
coherent thermal state~(\ref{two-modestate}).  We can see
that variance of different $Q_{ij}(r,\phi)$ gate transformed
number operators increase  with displacement.  While the
variance of $\mathcal{U}^\dagger(Q)\hat{N}\mathcal{U}(Q)$
with $e^r=\sqrt{2}$, $\phi= 0$, and $e^r=\sqrt{3}$, $\phi=0$
always remain higher than the variance of $\hat{N}$, the
variance of $\mathcal{U}^\dagger(Q)\hat{N}\mathcal{U}(Q)$
with $e^r=\sqrt{2}$, $\phi= \pi/2$ and $e^r=\sqrt{3}$,
$\phi=\pi/2$ crosses over the variance of $\hat{N}$ at a
certain value of the displacement parameter $u$.  Variance
of different $Q_{ij}(r,\phi)$ gate transformed number
operators as a function of squeezing $s$ is shown in
Fig.~\ref{fig:meanvar12}$(d)$.  As we can see, the squeezing
dependence of different variances exhibits a similar trend
as that of dependence on displacement.

Now we wish to relate the variances of transformed number
operators to variance of estimated Gaussian parameters. For
an $n$-mode system, quadrature operators $\hat{q}_i$ and
$\hat{p}_i$ can be expressed as 
\begin{eqnarray} \hat{q}_i
&=&  \hat{D}_i(1,0)^\dagger \hat{N} \hat{D}_i(1,0) - \hat{N}
-\frac{1}{2}, \nonumber \\ \hat{p}_i &=&
\hat{D}_i(0,1)^\dagger \hat{N} \hat{D}_i(0,1) - \hat{N}
-\frac{1}{2}.  \label{quadratur} 
\end{eqnarray} 
Averaging
the above equation yields Eq.~(\ref{meaneq}). Since
$\hat{N}$ and $\hat{D}_i(1,0)^\dagger \hat{N}
\hat{D}_i(1,0)$ are measured on different states, these
operators are uncorrelated and the expressions for the
variance of the quadratures become 
\begin{eqnarray}
\text{Var}(\hat{q}_i) &=&
\text{Var}(\hat{D}_i(1,0)^\dagger \hat{N} \hat{D}_i(1,0)) +
\text{Var}(\hat{N} ), \nonumber \\ \text{Var}(\hat{p}_i) &=&
\text{Var}(\hat{D}_i(0,1)^\dagger \hat{N} \hat{D}_i(0,1)) +
\text{Var}(\hat{N} ).  \label{quadrature} 
\end{eqnarray}
Thus, the variance of $\hat{q}_i$ quadrature, which
represents the quality of estimation of quadrature
$\hat{q}_i$, depends on both displacement $u$ and squeezing
$s$ as we can see from the above analysis. The optimization
of parameters $q_i$ and $p_i$ appearing in the displacement
gate $D_i(q_i,p_i)$ is required in order to minimize
$\text{Var}(\hat{q}_i)$.

Similarly we can express $\hat{q}_i^2$ as 
\begin{equation}
\hat{q}_i^2 =
6\bigg[\underbrace{\hat{\mathcal{U}}(P_i)^\dagger \hat{N}
\hat{\mathcal{U}}(P_i)}_{e^r=\sqrt{3},\phi=0}
-2\underbrace{\hat{\mathcal{U}}(P_i)^\dagger \hat{N}
\hat{\mathcal{U}}(P_i)}_{e^r=\sqrt{2},\phi=0} - \hat{N}
\bigg].  \end{equation} 
Thus the variance of $\hat{q}_i^2$
can be written as 
\begin{equation} \begin{aligned}
\text{Var}(\hat{q}_i^2) = 6\bigg[\underbrace{\text{Var}
(\hat{\mathcal{U}}(P_i)^\dagger \hat{N}
\hat{\mathcal{U}}(P_i))}_{e^r=\sqrt{3},\phi=0}
+&2\underbrace{\text{Var}(\hat{\mathcal{U}}(P_i)^\dagger
\hat{N} \hat{\mathcal{U}}(P_i))}_{e^r=\sqrt{2},\phi=0} \\ &+
\text{Var}(\hat{N}) \bigg].  \end{aligned} 
\end{equation}
We see from the above analysis that the variance of 
$\hat{q}_i^2$ also depends on both
displacement $u$ and squeezing $s$.
In this case too, a proper study of
the optimization of $P_i(r,\phi)$ gate parameters for the
minimization of $\text{Var}(\hat{q}_i^2)$ is needed.  Such
an analysis will be useful for the best estimation of Gaussian
state parameters. Similarly, various intra-mode correlation
terms such as $\text{Var}(\hat{p}_i^2)$ and
$\text{Var}(\hat{q}_i \hat{p}_i)$, as well as various inter-mode
correlation terms such as $\text{Var}(\hat{q}_i \hat{q}_j)$
and $\text{Var}(\hat{q}_i \hat{p}_j)$ can be expressed in
terms of the variances of different transformed number operators.
%%%%%%%%%%%%%%%%%%%%%%%%%%%%%%%%%%%%%%%%%%%%%%%%%%%%%%%%%%
\section{Concluding remarks} 
\label{sec:conclusion}
In this work we presented a Gaussian state tomography and
Gaussian process tomography scheme based on photon number
measurements. While the work builds upon the proposal given
in~\cite{rb-2015}, the current proposal offers an optimal
solution to the problem, with smaller number of optical
elements which renders the scheme more accessible to
experimentalists. After describing our optimal scheme for
Gaussian state tomography, we use it for estimation of a
Gaussian channel in an optimal way, where a total number of
$6n^2+n$ distinct measurements are required to determine
$6n^2+n$ parameters specifying a Gaussian channel.  Here we
have exploited the fact that $\text{Tr}(V)$ is the same for
all the output states corresponding to coherent state probes
with same or different mean.  Full state tomography of the
first coherent state probe yields an estimation of
$\text{Tr}(V)$ which can be used to estimate $\langle
\hat{N}\rangle$ for each of the remaining coherent state
probes, thus making the scheme optimal.  This in some sense
completes the problem of finding an optimal solution of the
Gaussian channel characterization posed in~\cite{rb-2015}.

It should be noted that our scheme is an improvement over
similar earlier schemes based on photon number measurements
and not over homodyne and heterodyne techniques which are
currently more prevalent. Similarly, the optimality is in
terms of the number of distinct experiments needed in the
scheme while each experiment will have to be repeated to obtain
the required average values. Having said so, it is worth
mentioning that  there have  been attempts to develop
homodyne measurement schemes using weak local oscillators
and PNRD~\cite{olivares-2017,walmsley-pra-2020} for use in
circumstances where strong local oscillators are not
desirable  and are essential for the
traditional homodyne scheme.  Our scheme based on PNRD
is an advancement in this direction as it requires no local
oscillator. Homodyne and heterodyne schemes go beyond
Gaussian states, whereas our present scheme is aimed only at
the estimation of Gaussian states. In principle, PNRD based
tomography schemes that go beyond Gaussian states can be
invented, however this aspect requires more investigation.
Finally, since PNRD measurements have become possible in
recent times, it is expected that in the coming years they
will become more practical and easier.

The analysis of variance in photon number measurements of
the original and transformed states shows that the variance
increases with the mean of the state and with the squeezing
parameter. Thus, this scheme is well suited for state with
small mean values or small displacements and small values of
squeezing. Extending the scheme for states with large mean
value but better estimation performance is under
consideration and will be reported  elsewhere.  While we
have chosen certain specific values of gate parameters (see
Eq.~(\ref{gate3})),  to extract information about the
parameters of the state, the effect of different values of
gate  parameters on the quality of estimates and
determination of optimal parameters that  maximize the
performance of the scheme needs further investigation.  The
optimality of the procedure may have a relationship with
mutually unbiased basis for the CV systems. Further analysis
of this aspect will require us to go beyond Gaussian states
and will be taken up elsewhere.
%%%%%%%%%%%%%%%%%%%%%%%%%%%%%%%%%%%%%%%%%%%%%%%%%%%%%%%%%%%
\section*{Acknowledgement}
R.S. acknowledges financial supports from {\sf SERB MATRICS
MTR/2017/000431} and {\sf
DST/ICPS/QuST/Theme-2/2019/General} Project number {\sf
Q-90}. Arvind acknowledges the financial
support from {\sf DST/ICPS/QuST/Theme-1/2019/General} Project
number {\sf Q-68}.
\end{document}